\documentstyle[prl,aps,twocolumn]{revtex}

\begin{document}

\title{
{\normalsize
\begin{flushright}
UB--ECM--PF 97/19\\
ITP--UH 23/97\\
hep-th/9709021\\[1.5ex]
\end{flushright}
}
Nonpolynomial gauge invariant interactions
of 1-form and 2-form gauge potentials
}
\author{
Friedemann Brandt\,$^a$, Norbert Dragon\,$^b$}

\address{
$^a$ Departament ECM,
Facultat de  F\'{\i}sica,
Universitat de Barcelona and
Institut de F\'{\i}sica
d'Altes Energies,
Diagonal 647,
E-08028 Barcelona, Spain.
E-mail: brandt@ecm.ub.es.
\\
$^b$ Institut f\"ur Theoretische Physik, Universit\"at Hannover,
Appelstr. 2, D-30167 Hannover, Germany.
E-mail: dragon@itp.uni-hannover.de.
\\[1.5ex]
\begin{minipage}{14cm}\rm\quad
A four dimensional gauge theory with nonpolynomial but local interactions
of 1-form and 2-form gauge potentials is constructed. The model is a 
nontrivial deformation of a free gauge theory with nonpolynomial dependence 
on the deformation parameter (= gauge coupling constant).
\\[2ex]
To appear in: Proceedings of the
31st International Symposium Ahrenshoop on the
Theory of Elementary Particles,
Buckow (Germany), September 2 - 6, 1997
\end{minipage}
}

\maketitle

Gauge invariant interactions of ordinary
gauge fields (= 1-form gauge potentials)
are very well known, the most famous one being
undoubtely the Yang--Mills interaction.
Much less is known about the possible gauge invariant
couplings between 1-form and 2-form gauge
potentials, although an important coupling of
that type is known for a long time: it is the
celebrated coupling of a 2-form gauge potential to
Chern--Simons forms which underlies
among others the Green--Schwarz anomaly
cancellation mechanism \cite{gs}.
Here we shall construct a rather different
interacting gauge theory for 1-form and 2-form gauge potentials
in four dimensions. As the model has local but
nonpolynomial interactions and gauge transformations,
its structure is in some respect more
reminiscent of gravitational interactions than
of Yang--Mills theory or couplings of Chern--Simons forms
to a 2-form gauge potential. It can however be formulated
in a polynomial first-order form, see \cite{HK} where the same
model was found by different means as a particular example in
a more general class of theories (one gets it from Eq. (17) of 
\cite{HK}, see remarks at the end of that paper).

Although we will not study supersymmetric field theories here, 
our construction was partly motivated by
the aim to gauge the ``central charge'' of the rigid N=2 supersymmetry
algebra realized on the so-called vector-tensor (VT) multiplet
\cite{sohnius} which arises naturally in string
compactifications \cite{deWit}. This problem was
investigated already in \cite{vt} and
could be relevant among others in order to classify the
still unknown couplings of the VT multiplet to N=2 supergravity.
Now, the central charge of the
VT multiplet is nothing but a bosonic
rigid symmetry of the standard (free) action for the
VT multiplet. Its name originates from the fact
that it occurs in the (anti)commutator of
two supersymmetry transformations with the same chirality.
This rigid
symmetry acts nontrivially only on the 1- and 2-form
gauge fields of the VT multiplet, as it is on-shell trivial
on the remaining component fields of the VT multiplet.
Therefore one can ask already in the
nonsupersymmetric case whether it
can be gauged in a reasonable way. This question is
interesting in its own right
and underlies our construction.

Our starting point is the standard free action for
two abelian 1-form gauge potentials $A=dx^\mu A_\mu$
and $W=dx^\mu W_\mu$ and a 2-form gauge potential
$B=(1/2)dx^\mu\wedge dx^\nu B_{\mu\nu}$ in flat four dimensional
spacetime. The Lagrangian reads
\begin{equation}
{\cal L}_0=-\frac 14\, (G_{\mu\nu}G^{\mu\nu}+F_{\mu\nu}F^{\mu\nu})
        -H_\mu H^\mu
\label{1}
\end{equation}
where
\begin{eqnarray}
G_{\mu\nu}&=&\partial_\mu W_\nu-\partial_\nu W_\mu\, ,
\nonumber\\
F_{\mu\nu}&=&\partial_\mu A_\nu-\partial_\nu A_\mu\, ,
\nonumber\\
H^\mu&=&\frac 12\,
\varepsilon^{\mu\nu\rho\sigma}\partial_\nu B_{\rho\sigma}\, .
\label{1a}
\end{eqnarray}
The action with Lagrangian (\ref{1}) has, among others,
a rigid symmetry generated by
\begin{equation}
\delta_z A_\mu=2H_\mu,\
\delta_z B_{\mu\nu}
       =\frac 12\,\varepsilon_{\mu\nu\rho\sigma}F^{\rho\sigma},\
\delta_z W_\mu=0.
\label{2}
\end{equation}
This rigid symmetry coincides indeed on-shell
with the central charge
of the N=2 supersymmetry algebra for the VT multiplet, cf.
\cite{sohnius}.

Our aim will now be to gauge the rigid symmetry (\ref{2}).
With this end in view, we look for appropriate extensions
$\Delta_z A_\mu$ and $\Delta_z B_{\mu\nu}$ of
$\delta_z A_\mu$ and $\delta_z B_{\mu\nu}$
transforming covariantly under
sought gauge transformations generated by
\begin{eqnarray}
\delta_\xi W_\mu &=&\partial_\mu \xi\, ,\nonumber\\
\delta_\xi A_\mu&=&g\,\xi\, \Delta_z A_\mu\, ,
\nonumber\\
\delta_\xi B_{\mu\nu}&=&g\,\xi\, \Delta_z B_{\mu\nu}
\label{7a}
\end{eqnarray}
where $\xi$ is an arbitrary field and
$g$ is a gauge coupling constant.
Following a standard receipe in gauge theories, we try
to covariantize partial derivatives of $A_\mu$ and
$B_{\mu\nu}$ by means of a covariant derivative
\begin{equation}
{\cal D}_\mu=\partial_\mu-g W_\mu \Delta_z
\label{3}
\end{equation}
where $\Delta_z$ is
the sought extension of $\delta_z$. We now try to
covariantize (\ref{2}) by replacing there $\delta_z A_\mu$ and
$\delta_z B_{\mu\nu}$ with $\Delta_z A_\mu$ and $\Delta_z B_{\mu\nu}$
respectively, and $\partial_\mu$ with ${\cal D}_\mu$.
Explicitly this yields
\begin{eqnarray}
\Delta_z A_\mu&=&\varepsilon_{\mu\nu\rho\sigma}
      (\partial^\nu B^{\rho\sigma}-gW^\nu\Delta_z B^{\rho\sigma}),
\nonumber\\
\Delta_z B_{\mu\nu}&=&\varepsilon_{\mu\nu\rho\sigma}
                 (\partial^\rho A^\sigma-gW^\rho\Delta_z A^\sigma).
\label{4}
\end{eqnarray}
(\ref{4}) determines
$\Delta_z A_\mu$ and $\Delta_z B_{\mu\nu}$. Indeed, inserting
the second equation (\ref{4}) in the first one, we get
an equation for $\Delta_z A_\mu$,
\begin{equation}
(E\delta_\mu^\nu+2g^2W_\mu W^\nu)\Delta_z A_\nu=2Z_\mu\, ,
\label{4b}
\end{equation}
where
\begin{equation}
E=1-2g^2W_\mu W^\mu,\ 
Z_\mu=H_\mu+gF_{\mu\nu}W^\nu.
\label{4c}
\end{equation}
To solve (\ref{4b}) for $\Delta_z A_\nu$, we only need to
invert the matrix
$E\delta_\mu^\nu+2g^2W_\mu W^\nu$. The inverse is
\begin{equation}
{V_\mu}^\nu=E^{-1}(\delta_\mu^\nu-2g^2W_\mu W^\nu) .
\label{4f}
\end{equation}
(\ref{4b}) and (\ref{4}) yield now
\begin{equation}
\Delta_z A_\mu=2{\cal H}_\mu\ ,\
\Delta_z B_{\mu\nu}
=\frac 12\,\varepsilon_{\mu\nu\rho\sigma}{\cal F}^{\rho\sigma}
\label{5}
\end{equation}
where
\begin{eqnarray}
{\cal H}_\mu&=&E^{-1} (Z_\mu-2g^2 W_\mu W_\nu H^\nu),
\nonumber\\
{\cal F}_{\mu\nu}&=&F_{\mu\nu}-4gE^{-1} W_{[\mu}Z_{\nu]}\, .
\label{6}
\end{eqnarray}

Recall that our goal was to find gauge transformations
(\ref{7a}) under which $\Delta_z A_\mu$ and $\Delta_z B_{\mu\nu}$
transform covariantly. We can now examine whether
we have reached this goal. This amounts to check
whether ${\cal H}_\mu$ and ${\cal F}_{\mu\nu}$ transform
covariantly (i.e., without derivatives of
$\xi$) under the gauge transformations
\begin{eqnarray}
\delta_\xi A_\mu&=&2g\,\xi\, {\cal H}_\mu\ ,
\nonumber\\
\delta_\xi B_{\mu\nu}&=&
\frac 12\,g\,\xi\,\varepsilon_{\mu\nu\rho\sigma}{\cal F}^{\rho\sigma},
\nonumber\\
\delta_\xi W_\mu&=&\partial_\mu \xi\, .
\label{7}
\end{eqnarray}
The answer is affirmative, i.e.\ neither
$\delta_\xi{\cal H}_\mu$ nor $\delta_\xi{\cal F}_{\mu\nu}$
contain derivatives of $\xi$. Indeed, an elementary, though somewhat 
lengthy calculation yields
\begin{equation}
\delta_\xi {\cal H}_\mu=g\,\xi\,\Delta_z{\cal H}_\mu\, ,\
\delta_\xi{\cal F}_{\mu\nu}=g\,\xi\,\Delta_z {\cal F}_{\mu\nu}
\label{8}
\end{equation}
with
\begin{eqnarray}
\Delta_z{\cal H}_\mu&=&
{V_\mu}^\nu (\partial^\rho{\cal F}_{\rho\nu}
-4gW^\rho\partial_{[\rho}{\cal H}_{\nu]}),
\nonumber\\
\Delta_z {\cal F}_{\mu\nu}&=&4
(\partial_{[\mu}{\cal H}_{\nu]}
-gW_{[\mu}\Delta_z{\cal H}_{\nu]}).
\label{8a}
\end{eqnarray}

To construct an action which is invariant under
the gauge transformations (\ref{7}), it is helpful
to realize that the transformations (\ref{8a}) are
nothing but
\begin{equation}
\Delta_z {\cal H}_\mu={\cal D}^\nu{\cal F}_{\nu\mu},\
\Delta_z{\cal F}_{\mu\nu}=4{\cal D}_{[\mu}{\cal H}_{\nu]}
\label{8b}
\end{equation}
where the second identity is obvious from
(\ref{8a}), whereas the verification of the first one
is slightly more involved. Combining (\ref{8}) and
(\ref{8b}) it is now easy to verify that
\begin{eqnarray*}
\lefteqn{
\delta_\xi\left(
-\frac 14\, {\cal F}_{\mu\nu}{\cal F}^{\mu\nu}
-{\cal H}_\mu {\cal H}^\mu\right)}
\\
&=&2g\,\xi\,{\cal D}_\nu({\cal H}_\mu {\cal F}^{\mu\nu})
\\
&=&2g\,\xi\,\partial_\nu({\cal H}_\mu {\cal F}^{\mu\nu})
-2g^2\xi\, W_\nu\Delta_z({\cal H}_\mu {\cal F}^{\mu\nu})
\\
&=&\partial_\nu(2g\,\xi\, {\cal H}_\mu {\cal F}^{\mu\nu})
   -\delta_\xi(2gW_\nu{\cal H}_\mu {\cal F}^{\mu\nu}).
\end{eqnarray*}
This implies immediately that the Lagrangian
\begin{eqnarray}
{\cal L}&=&
-\frac 14\, (G_{\mu\nu}G^{\mu\nu}+{\cal F}_{\mu\nu}{\cal F}^{\mu\nu})
\nonumber\\
&& -{\cal H}_\mu {\cal H}^\mu-2gW_\mu{\cal H}_\nu {\cal F}^{\mu\nu}
\label{10}
\end{eqnarray}
transforms under $\delta_\xi$
into a total derivative,
\begin{equation}
\delta_\xi{\cal L}=
\partial_\nu(2g\,\xi\, {\cal H}_\mu {\cal F}^{\mu\nu}).
\end{equation}
Hence, the action with Lagrangian (\ref{10}) is gauge
invariant under $\delta_\xi$.
Evidently it is also invariant under
the following standard gauge transformations acting only on
$A_\mu$ and $B_{\mu\nu}$ respectively:
\begin{equation}
\delta_\Lambda A_\mu=\partial_\mu \Lambda,\
\delta_\lambda B_{\mu\nu}=\partial_{[\mu}\lambda_{\nu]}\, .
\label{othersymm}
\end{equation}
Inserting finally the explicit expressions (\ref{6})
in (\ref{10}), the Lagrangian reads
\begin{eqnarray}
{\cal L}&=&
-\frac 14\, (G_{\mu\nu}G^{\mu\nu}+F_{\mu\nu}F^{\mu\nu})
\nonumber\\
&& -E^{-1} Z_\mu Z^\mu+2E^{-1}(gW_\mu H^\mu)^2
\label{11}
\end{eqnarray}
with $E$ and $Z_\mu$ as in (\ref{4c}).

It is now easy to compute
the Euler--Lagrange derivatives of ${\cal L}$ with
respect to the fields. The result is
\begin{eqnarray}
\frac{\hat \partial{\cal L}}{\hat \partial A_\mu}&=&
\partial_\nu {\cal F}^{\nu\mu},
\label{eomA}\\
\frac{\hat \partial{\cal L}}{\hat \partial B_{\mu\nu}}&=&
-\varepsilon^{\mu\nu\rho\sigma}\partial_\rho {\cal H}_\sigma\, ,
\label{eomB}\\
\frac{\hat \partial{\cal L}}{\hat \partial W_\mu}&=&
\partial_\nu G^{\nu\mu}+2g {\cal F}^{\mu\nu}{\cal H}_\nu\, .
\label{eomW}
\end{eqnarray}
Note that the equations of motion for $A_\mu$ and $B_{\mu\nu}$
obtained from (\ref{eomA}) and (\ref{eomB}) are not covariant
under $\delta_\xi$,
in contrast to the equation of motion for $W_\mu$ following
from (\ref{eomW}). However, from (\ref{8a}) and (\ref{8b}) it
is obvious that (\ref{eomA}) and (\ref{eomB})
can be combined to covariant expressions too,
which illustrates once again a general property of the
equations of motion in gauge theories \cite{ten}.
The covariant form of the equations of motion reads
\begin{eqnarray*}
& {\cal D}_\nu {\cal F}^{\mu\nu}=0, &\\
& {\cal D}_{[\mu}{\cal H}_{\nu]}=0, &\\
& \partial_\nu G^{\nu\mu}+2g {\cal F}^{\mu\nu}{\cal H}_\nu=0.&
\end{eqnarray*}

To summarize, we have constructed an
interacting four dimensional gauge theory with
Lagrangian (\ref{10}) resp. (\ref{11})
for two ordinary gauge fields $A_\mu$ and $W_\mu$
and an antisymmetric gauge field
$B_{\mu\nu}$. The key feature of this gauge theory
is its gauge invariance under the transformations
(\ref{7}) which gauge the rigid
symmetry (\ref{2}) of the free action with Lagrangian
(\ref{1}). Both the Lagrangian
and the gauge transformations (\ref{7}) are
nonpolynomial in the gauge coupling
constant $g$ and the gauge field $W_\mu$. Nevertheless
they are local, for the Lagrangian and the gauge transformations
are still quadratic and linear in derivatives respectively.
Note that the Lagrangian and the gauge transformations constitute
a consistent deformation of the free Lagrangian (\ref{1}) and its gauge
symmetries in the sense of \cite{bh}.
In particular one recovers the free theory and
its gauge symmetries for $g=0$.
The generalization of all above formulas to curved
spacetime is obvious.

Let us finally compare our results to those
of \cite{vt} where the central charge of
the VT multiplet was gauged.
First we have presented the action
and the gauge transformations
in an explicit and manifestly local form.
In contrast, in \cite{vt} both the action and
the gauge transformations are only implicitly defined
(the formulas given in \cite{vt} result in local expressions
too \cite{PP}). Furthermore
our results appear to differ from those
of \cite{vt}, even when the latter are restricted to the
particular nonsupersymmetric
case studied here. In particular, neither
the Lagrangian (\ref{11}) nor the gauge transformations
(\ref{7}) contain Chern--Simons
terms of the type occurring in \cite{vt}. This might
signal that such terms are actually not needed in order
to gauge the central charge of the VT multiplet.
Of course, in contrast to \cite{vt}, we did not study the
supersymmetric case, and therefore we cannot clarify this issue here.
\medskip

\noindent{\em Acknowledgement:}

We thank Piet Claus, Bernard de Wit, Michael Faux,
Marc Henneaux and Piet Termonia
for useful discussions and comments.
F.B. was supported by the Spanish ministry of education and
science (MEC).

\end{document}